\documentclass[conference]{IEEEtran}
\IEEEoverridecommandlockouts
\usepackage{cite}
\usepackage{float}
\usepackage{amsmath,amssymb,amsfonts}
\usepackage{algorithm}
\usepackage{algpseudocode}
\usepackage[pagebackref=true,breaklinks=true,colorlinks,bookmarks=false]{hyperref} 
\usepackage[table]{xcolor}
\usepackage{colortbl}
\usepackage{subcaption}
\usepackage{times}
\usepackage{makecell}
\usepackage{multirow}
\usepackage{url}            
\usepackage{booktabs}       
\usepackage{amsfonts}       
\usepackage{nicefrac}       
\usepackage{microtype}      
\usepackage{amsmath}
\usepackage{bbding}
\usepackage{float}
\usepackage{caption}
\usepackage{subcaption}
\usepackage{wrapfig}

\usepackage{graphicx}
\usepackage{textcomp}
\usepackage{xcolor}
\usepackage{listings}
\usepackage{adjustbox}
\def\BibTeX{{\rm B\kern-.05em{\sc i\kern-.025em b}\kern-.08em
    T\kern-.1667em\lower.7ex\hbox{E}\kern-.125emX}}
\begin{document}

\title{Flover: A Temporal Fusion Framework for Efficient Autoregressive Model Parallel Inference  \\
}

\author{
    \IEEEauthorblockN{
        Jinghan Yao,
        Nawras Alnaasan,
        Tian Chen,
        Aamir Shafi,
        Hari Subramoni,
        Dhabaleswar K. (DK) Panda
    }
    \IEEEauthorblockA{
        \textit{Department of Computer Science and Engineering} \\
        \textit{The Ohio State University}\\
        Columbus, OH, U.S. \\
        \{yao.877, alnaasan.1, chen.9891, shafi.16, subramoni.1\}@osu.edu, panda@cse.ohio-state.edu
    }
}

\maketitle
\thispagestyle{plain}
\pagestyle{plain}

\begin{abstract}

Autoregressive models, despite their commendable performance in a myriad of generative tasks, face challenges stemming from their inherently sequential structure. Inference on these models, by design, harnesses a temporal dependency, where the current token's probability distribution is conditioned on preceding tokens. This inherent characteristic severely impedes computational efficiency during inference as a typical inference request can require more than thousands of tokens, where generating each token requires a load of entire model weights, making the inference more memory-bound. The large overhead becomes profound in real deployment where requests arrive randomly, necessitating various generation lengths. Existing solutions, such as dynamic batching and concurrent instances, introduce significant response delays and bandwidth contention, falling short of achieving optimal latency and throughput. To address these shortcomings, we propose Flover -- a temporal fusion framework for efficiently inferring multiple requests in parallel. We deconstruct the general generation pipeline into pre-processing and token generation, and equip the framework with a dedicated work scheduler for fusing the generation process temporally across all requests. By orchestrating the token-level parallelism, Flover exhibits optimal hardware efficiency and significantly spares the system resources. By further employing a fast buffer reordering algorithm that allows memory eviction of finished tasks, it brings over $11\times$ inference speedup on GPT and $16\times$ on LLAMA compared to the cutting-edge solutions provided by NVIDIA FasterTransformer. Crucially, by leveraging the advanced tensor parallel technique, Flover proves efficacious across diverse computational landscapes, from single-GPU setups to distributed scenarios, thereby offering robust performance optimization that adapts to variable use cases.
\end{abstract}


\section{Introduction}
\label{intro}

Large-scale artificial intelligence (AI) models, especially autoregressive ones, are helping make significant strides in several important areas such as Natural Language Processing (NLP), time-series forecasting, and signal processing. Autoregressive models, including notable Large Language Models (LLMs) like the Generative Pretrained Transformer (GPT) series~\cite{brown2020language,gpt-j,https://doi.org/10.48550/arxiv.2204.06745,openai2023gpt4,radford2019language, gpt-neo, touvron2023llama}, stand out for their ability to predict successive outputs based on preceding ones and the entire input sequence. This inherent characteristic of forming temporal dependencies among outputs is a characteristic that is particularly pronounced in autoregressive models.

The training of these autoregressive models is a computationally demanding process due to the sheer volume of parameters involved, the extensive sequence lengths, and the requirement of techniques such as beam search and top-k sampling. However, it's important to note that training is largely a one-time effort, often done in-house before the model is made available to the public. A technique known as {\em sequence masking} for parallelization has proved instrumental in mitigating this challenge. By leveraging the available ground truth for all output sequences in the training dataset, sequence masking enables the simultaneous processing of different parts of an input sequence, thereby considerably accelerating the training process.

While the optimization of the training phase is crucial, the real-time user experience predominantly hinges on the efficiency of the inference phase. This phase, however, encounters unique challenges due to the much smaller batch size per request, random flow of requests, and various inference configurations. Along with the sequential dependency, traditional parallel inference strategies such as dynamic batching and concurrent instances find huge gaps in accelerating the inference service on servers. Therefore, while the training phase can be expedited via sequence masking and dedicated tricks, optimizing the inference phase, which directly impacts user experience, requires a more tailored approach.


\subsection{Problem Statement}
\label{sec:intro:stmt}

\begin{figure*}[t]\centering
\includegraphics[width=0.8\linewidth]{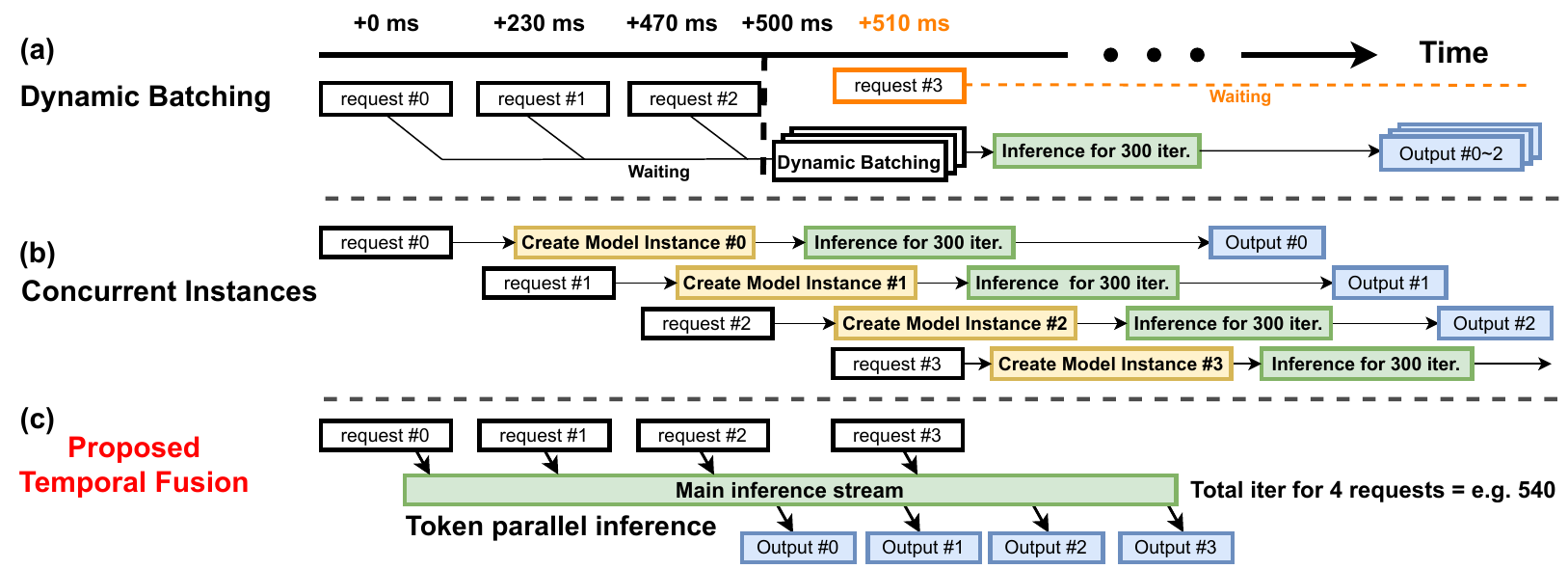}
\caption{Workflow comparison on dynamic batching, concurrent model instances, and our Temporal fusion. Time stamps on the line give an example of different arrival times of requests. For dynamic batching, we assume the time window is 500ms, though this may vary in real cases. In this example, each inference request asks for 300 iterations.}
\label{fig:Structure}
\end{figure*}

With the rapid advancement of AI, inference servers routinely grapple with the processing of multiple concurrent inference requests from autoregressive models. Current methodologies such as dynamic batching and concurrent model instances, employed by inference frameworks like Microsoft DeepSpeed~\cite{rasley2020deepspeed,aminabadi2022deepspeed} and NVIDIA Triton Inference Server~\cite{Triton}, become much less effective when confronted with autoregressive models, as they either significantly delay the inference to a heuristic time window or launching too many model instances causing severe hardware contention, the degradation in performance exacerbates rapidly as the number of parallel requests increase, as shown in our thorough profiling in~\ref{profiling}. Fig~\ref{fig:Structure} provides a real deployment case to compare these methods and our proposed temporal fusion strategy, a detailed discussion will be provided in~\ref{limitation}.

\subsection{Motivation}
\label{sec:intro:motivation}
A pressing need in today's AI landscape is the development of robust strategies capable of parallelizing these temporally dependent inference requests. This would effectively enhance system efficiency, improve throughput, and reduce response time, ultimately leading to a better user experience and wider applicability of these advanced AI models. In autoregressive structure, the inherent constraints impede the potential of parallel processing any single request, though, when dealing with multiple of them, we indeed can adopt a parallel pipelining method across all requests. In fact, it is more common and particularly prominent in real-time applications and scenarios where the inference server constantly receives new requests as it is processing previous ones. As existing solutions become ineffectual in handling such scenarios, there is an urgent need to enhance the efficiency of the inference process in autoregressive models, a necessity recognized by the AI community.
And this focus is paving the way for the next wave of advancements in AI, aimed at making these powerful models more accessible and efficient.


\subsection{Contributions}
\label{sec:intro:contrib}
In this work, we propose \textbf{Flover}, a temporal fusion framework tailored to the context of parallel inference in autoregressive models. The main contribution of Flover is to promptly process incoming requests, eliminating the need for batching or time window allocation, while not triggering the launch of redundant model instances or kernel calls. 

Flover only maintains one main computing stream to handle any number of requests throughout the lifecycle of inference, largely reducing the overhead in numerous separated kernel calls and scheduling redundant collective communicators.


The paper makes the following contributions: 


\begin{enumerate}
    \item We introduce a novel {\em temporal fusion framework} for propelling autoregressive model inference by leveraging the temporality property of inferring generative tasks, delivering superior and more fine-grained parallelism beyond all current solutions.
    \item We thoroughly analyze multiple real inference scenarios and compare our solution with the cutting-edge NVIDIA FasterTransformer~\cite{Triton, ft}, where Flover provides speedups of \textbf{11x} on GPT-J~\cite{gpt-j} 6B and Llama~\cite{touvron2023llama} 7B, \textbf{7x} on Llama~\cite{touvron2023llama} 13B, \textbf{13x} on Llama~\cite{touvron2023llama} 33B, and \textbf{16x} on Llama~\cite{touvron2023llama} 65B models.
    \item We design the Flover scheduler to significantly reduce hardware resource requirements by saving $93.9\%$ of CPU threads and $96.9\%$ of GPU kernel launching compared to FasterTransformer~\cite{Triton, ft}, in parallel inference 32 requests.
    \item We design an efficient memory shuffle algorithm that can reorder requests' buffers, such that significantly reduce computing workload and communication message volume, further providing $23\%$ faster inference.
    \item To the best of our knowledge, Flover provides a high-performance solution in accelerating autoregressive model inference with full support on various token generation configurations. It is not restricted to hardware resources, delivering the above performance gain not only on single GPU inference, and also seamlessly works with the advanced tensor parallel~\cite{shoeybi2019megatron} technique to accelerate distributed inference.
\end{enumerate}
The code for our project is open-sourced and available online.\footnote{\url{https://github.com/YJHMITWEB/Flover.git}}


\section{Background}

\begin{table*}[ht] \centering
\begin{tabular}{c| c | c |c |c |c 
}
\toprule
    & {Variable generation}& Requests fused    & Buffer  &  \multicolumn{2}{c}{Parallel inference 32 requests} \\
    \cline{5-6}
   & {length}& processing   &  reordering &{\# of threads required} & {\# of kernel launch} \\ 
\midrule
FasterTransformer~\cite{Triton, ft} & \checkmark &     &   & 32 (+1)         & 32   \\
ORCA~\cite{yu2022orca} &   & \checkmark &     & /     & /          \\
\textbf{Flover (ours)} & \checkmark  & \textbf{\checkmark} & \textbf{\checkmark} & \textbf{1} (+1)     & \textbf{1}          \\
\bottomrule
\end{tabular}
\caption{Summarizing  state-of-the-art parallel inference frameworks.}
\label{tab:ft_dp_oc}
\end{table*}

\subsection{Autoregressive models} Deep learning architectures comprise a variety of models with unique traits and applications. Broadly, these models fall into two categories: non-autoregressive and autoregressive, differentiated by their operational patterns.

Non-autoregressive models, such as ResNet~\cite{he2016deep}, Vision Transformer~\cite{dosovitskiy2020image, lu2021soft} for image classification, and YOLO~\cite{redmon2016you} for object detection, function as feed-forward networks. These models independently process each input through a sequence of transformations, thereby producing output in a single pass through the model. 

On the other hand, autoregressive models, such as GPT and Llama models used in language modeling tasks, represent a different operational paradigm, distinguished by their temporal dependencies. Unlike non-autoregressive models, these models necessitate multiple passes through the model, using the output from one run as the input for the next. This inherently sequential pattern is critical in tasks such as natural language processing and time series analysis, where maintaining the order of data points is vital. In particular, this sequential nature is a crucial characteristic of language models like GPT, enabling them to generate coherent and contextually appropriate language by taking into account the previous tokens. 

\subsection{Parallel inference frameworks}
Parallel inference frameworks, most notably FasterTransformer, play a critical role in boosting the efficiency of Transformer-based models. The FasterTransformer library by NVIDIA has been meticulously designed to optimize and expedite the execution of such models, including Transformer-based autoregressive architectures such as the GPT series. The library incorporates a set of optimization strategies such as kernel fusion and the use of half-precision (FP16) operations, a method particularly suited for deep learning computations. The broad compatibility of FasterTransformer with popular deep learning frameworks like TensorFlow and PyTorch further consolidates its widespread utility in the realm of Transformer-based applications.

\section{Challenges and Limitations of Existing Approaches}

\label{limitation}

In the quest for efficient inference, general solutions such as {\textbf{dynamic batching} and \textbf{concurrent model instances} as shown in Fig~\ref{fig:Structure} have been integrated into frameworks like Microsoft DeepSpeed~\cite{rasley2020deepspeed,aminabadi2022deepspeed} and NVIDIA Triton Inference Server~\cite{Triton}. Table~\ref{tab:ft_dp_oc} compares the cutting-edge inference frameworks with our proposed Flover design. We first analyze the common strategies of parallel inference, a detailed stats will be provided in~\ref{profiling}.

\textbf{Dynamic batching} allows the server to wait within a time window $\tau$, which is pre-defined according to the estimated volume of requests. Requests that arrive within the $i_{th}$ time window $\tau_i$ will be packed together along the batch dimension. When the time window is reached or the maximum requests are presented, the packed batch $b_i$ will be passed into the inference model as a whole for more efficient processing. Since in inference scenarios, a single request usually has a much smaller batch size compared to training, packing requests to a larger batch will lead to higher GPU utilization and throughput. Though, determining the time window can be heuristic and exhibits no flexibility. For example, in Fig~\ref{fig:Structure} (a), request \#0 that arrived at the beginning of a time window will have to wait for the whole window until it can be processed, this could lead to severe overhead in latency and also prevent possible overlap of computation. Even worse, as shown in Fig~\ref{fig:Structure} (a), request 3 arrives at 510 ms, thus it has to wait until the currently running batch finishes. In autoregressive models, this will significantly increase the response time.

\textbf{Concurrent model instances} allows the immediate launching of a new inference instance once a request arrives, and the instance will only infer this request. As shown in Fig~\ref{fig:Structure} (b), the inference server first loads the model weight into the GPU memory. Then, for each request it receives, a new thread will be spawned by the server and it will create a new instance of the inference model. As more and more requests arrive, the server will continuously spawn new threads to handle each of them separately. Notice that all instances will share the same model weight that was pre-loaded in the global memory, so that the overall memory consumption is still reasonable. However, this method can introduce severe overhead because each model instance can consume a massive amount of memory bandwidth during computing, and when multiple instances run concurrently, they compete for the same resources, draining the bandwidth, causing frequent context switching on both CPU and GPU sides, therefore creating a resource contention scenario, leading to severe performance degradation.

\begin{figure*}[t]
\centering
\includegraphics[width=1.0\linewidth]{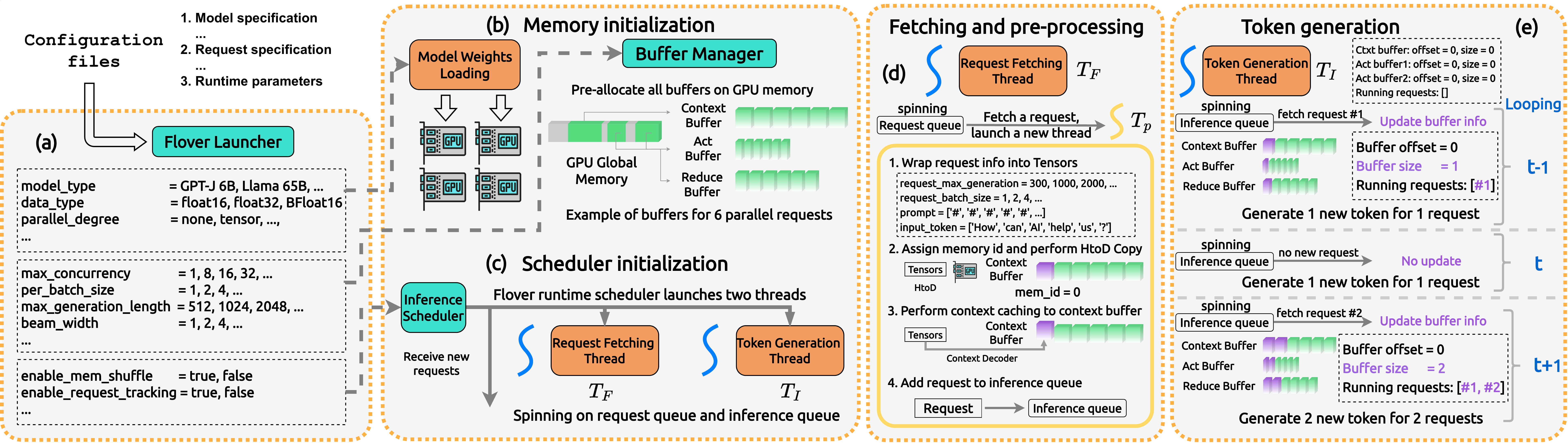}
\caption{Schematic illustration of the proposed Temporal Fusion Framework for autoregressive models. (a) Flover launcher reads configurations, (b) memory initialization on both model weights loading and buffer pre-allocation, (c) Inference scheduler is initialized and spawns two threads for fetching requests and token generation, (d) Fetching thread spins on the request queue and perform pre-processing, (e) token generation thread spins on inferene queue and generates tokens in parallel for all requests.
}
\vspace{-1em}
\label{fig:pipeline}
\end{figure*}
ORCA~\cite{yu2022orca} is an autoregressive model inference framework that aims to tackle the resource contention situation by applying a selection engine to perform the batching of requests. However, their design requires determining the batching strategy at every iteration and as shown in Table~\ref{tab:ft_dp_oc}, it can only handle requests that are uniform in generation length, which is less likely to be true in real deployment. Also as it is not public accessible, preventing us from further analysis.


\section{Preliminaries}
\label{method}

To schematically demonstrate our method, let's first define what a request is in autoregressive model inference. Consider the GPT models, a request $R_i$ has the following domains:
{\small 
\begin{itemize}
    \item $R_i$
    \begin{itemize}
        \item Batch size: A positive integer $n$, e.g. 1
        \item Input words: $n$ lists of words, 
        
        e.g. [`How', `can', `AI', `help', `humans', `?']
        \item Max Generation Length: A positive integer, e.g. 300
    \end{itemize}
\end{itemize}
}

The above request indicates that for such a question ``How can AI help humans?", the inference server is allowed to generate a response of at most 300 words. According to the model specification, the inference process might terminate early if it outputs an EOS word, such as ``\$", denoting the completion of the answer. Or, if it reaches the maximum length (e.g. 300), it will force the inference process to stop. 

Next, we will analyze two real inference scenarios where requests' arrival follows a constant time interval $\tau$ or the Poisson process. The memorylessness property of the Poisson process aligns with the nature of independent request arrivals, while the burstiness and sparsity observed in deep learning systems can be accommodated within this paradigm. Considering the arrival of requests conforms to $P(k) = \frac{e^{-\lambda}\lambda^k}{k!}$, the arrivals of requests occur randomly and independently over time. $\lambda$ denotes the expected number of arrivals that occur in a unit interval of time, and $P(k)$ represents the probability of $k$ requests arriving within a unit time interval. Then the time interval $x$ between two arrivals can be modeled by Exponential distribution $f(x)=\lambda e^{-\lambda x}, x \geq 0$.

Utilizing both paradigms enables us to gain insights into the request arrival patterns, facilitating efficient resource allocation and capacity planning within our design.

\section{Framework design} 
With all the insights we have, we propose Flover, a temporal fusion framework for propelling inference on autoregressive models. First, we make the following clarifications. For every request, we decompose its inference pipeline into two main phases, namely, 1). Pre-processing 2). Token generation.

Fig~\ref{fig:Structure} (c) shows the abstract workflow of Flover, and Fig~\ref{fig:pipeline} further shows more details. We illustrate the Flover framework by it's components, namely, (a) Flover Launcher, (b) Buffer Manager, and (c) Inference Scheduler.



\subsection{Memory and scheduler initialization}
\label{ctx:initialization}
Flover Launcher in Fig~\ref{fig:pipeline} (a) will first read user-provided configuration files, in which the type of the model, inference data type, and parallel strategy are defined. As the information is sufficient, it will immediately issue the model launching by loading the model weights from the disk to GPU memory. 

Next, Flover will read the meta specification of requests which will be conformed to by all requests in this launch. Given the actual scenarios where the frequency and intensity of inference requests are pre-determined, \texttt{max\_concurrency} sets the maximum number of requests that can be concurrently processed. Notice that the theoretical maximum of this value is directly related to available GPU memory. In our experiments, we set it up to 32 as it is reaches the memory limit of an NVIDIA A100 GPU, denoting that there could be at most 32 requests running in the inference stream, in other words, 32 tokens will be generated per iteration. More requests will be waiting until a running one finishes. Considering it with other parameters such as the batch size, the max generation length, and the width of beam search, Buffer Manager in Fig~\ref{fig:pipeline} (b) could calculate the upper bound of memory usage during the running. Notice that Flover applies a pinned and reusable memory management, which means that every buffer used by the framework will be pre-allocated, avoiding any dynamic allocation during inference. Here are some advantages: 1) Parallel inference frameworks are considered to last long on the server, meaning that as long as the workload and flow are reasonably stable, we could expect all memory that it pre-allocates will be in use without any waste. 2) Allocating memory for all requests will guarantee their buffers are contiguous and adjacent to each other, which could largely increase the memory utilization and benefit the buffer reordering technique which we will introduce later in~\ref{shuffle}, as it further brings about $23\%$ of improvement on inference latency. Fig~\ref{fig:pipeline} (b) shows the pre-allocation for three different buffers that will be used in each round of generation. Note that in the real model, the number of different buffers used in inference is much more, and each buffer varies in size.

After GPU buffer is all set, we launch the inference scheduler, preparing to fetch requests and perform token generation. Flover adopts a queue-spinning strategy for handling new requests. Specifically, the scheduler will create two queues, a request queue~$Q_R$, and an inference queue~$Q_I$. And it launches two threads, a request fetching thread~$T_F$ and a token generation thread~$T_I$. During the whole running time of Flover, $T_F$ will spin on $Q_R$, checking if there are any new requests received by the server. Once it gets a new request from $Q_R$, it will spawn a new thread~$T_P$ (denoted as the yellow string in Fig~\ref{fig:pipeline} (d)) for pre-processing this request. 

\subsection{Fetching and pre-processing}
\begin{algorithm}[ht]
    \caption{Initialization, fetching requests, and performing pre-processing in parallel}
    \label{alg:buffer}
    \scriptsize
    
    \begingroup
    \color{teal}
    /* Initiate mutex queues and buffers */ \\
    \endgroup 
    \texttt{RequestQueue reque;} \\
    \texttt{InferenceQueue inque;} \\
    
    \begingroup
    \color{teal}
    /* Pre-allocate buffers */ \\
    \endgroup 
    DeviceMalloc(\texttt{ctxt\_buffer, ln\_buffer, act\_buffer, reduce\_buffer}) \\

    \begingroup
    \color{magenta}
    \textbf{{While not finish}} /* Loop, keep pulling for new requests */ \\
    \endgroup
    \hspace*{2em}\textbf{\textcolor{violet}{If}} {\texttt{reque.get(req)}} \textbf{\textcolor{violet}{then}} \textcolor{blue}{\texttt{std::thread()}}\\
    \begingroup
    \color{teal}
    \hspace*{4em}/* Fetching a request and assigning its memory index */ \\
    \endgroup 
    \hspace*{4em}\texttt{offset = AssignMemId() * per\_request\_size;} \\
    
    \begingroup
    \color{teal}
    \hspace*{4em}/* Start computing context cache on corresponding buffer */ \\
    \endgroup 
    \hspace*{4em}\texttt{preprocessing->run(ctxt\_buffer+offset);} \\
    
    \begingroup
    \color{teal}
    \hspace*{4em}/* Current request is ready for token generation */ \\
    \endgroup 
    \hspace*{4em}\texttt{inque.add(req);} \\
    \begingroup
    \color{violet}
    \hspace*{2em}\textbf{\textcolor{violet}{End If}} \\
    \endgroup 
    \begingroup
    \color{magenta}
    /* End loop */ \\
    \endgroup 
\vspace{-1em}
\end{algorithm}
As we mentioned previously, that a new request coming from the user only contains the raw input data, here we need to wrap the data into a more compact data structure, e.g. tensors or tensor maps, and we need to copy from host memory to GPU buffer, as shown in Fig~\ref{fig:pipeline} (d). It is noteworthy that the Memcpy is also performed on pre-allocated buffers as it will later be used by the model, therefore, at this step, we have to decide which slot of the buffer should the data go. In Fig~\ref{fig:pipeline}, for better visualization, we draw 3 buffers, each with 6 slots, meaning that we allow a maximum of 6 requests' concurrent inference. Namely, they are context buffer, act buffer 1, and act buffer 2. Assuming that the new request is assigned with a unique memory index \texttt{mem\_{id}=0}, the following context decoder will then process this request and put the intermediate result to slot 0 of context buffer. At this step, the context of the request has been cached in memory, which means it is ready for token generation. As shown in Algorithm~\ref{alg:buffer}, for simple clarification, we list the basic outline of its logic.

Before proceeding to the next section, we would like to further clarify the concurrent context caching design. When a new request arrives, it might be as simple as the case we introduced above, which only requires data packing before going to the model, but in more often cases, the pre-processing could include many more customized operations such as keyword filtering, sensitivity checking, spell correction, or a heavier context decoder, which are not parallelizable. In order to avoid any congestion, we shortly spawn new threads to handle the potential high concurrency and they no longer exist after this step. We emphasize that for the typical autoregressive generation task, the pre-processing part takes over less than $1\%$ of the overall elapsed time, therefore the potential of draining the system resource from multi-thread subscription is negligible.

\subsection{Parallel token generation for multiple requests} 

In generative models, token generation is the most time-consuming process. A request that asks for a 1000 words response will simply cost twice the time of a 500 words request, due to the unbreakable data dependency existing in the \texttt{for} loop. Fortunately, generating tokens for different requests follows identical procedures. As shown in Fig~\ref{fig:pipeline} (e), suppose that we have two requests, where the second request joins the inference stream at iteration t+1, while the first request has been in inference since iteration t-1. In every iteration, the model performs completely identical operations, for example, cuBLAS kernel calls, Layernorm, Allreduce, Allgather, and the only difference is in the value of data itself. This means that instead of inference two requests separately with multiple kernel calls or collective operations, we can simply put their buffers adjacent in memory and launch a single kernel call to perform the computation and communication. Therefore, at iteration t-1 and t, Flover is generating 1 token for the first request. At iteration t+1, since the second request is received by the generation thread, it will update the corresponding buffer size to 2, so that all following kernel operations can directly work on these buffers.

Algorithm~\ref{alg:fusion} and Fig~\ref{fig:memory_shuffle1} (first row) illustrate how this temporal fusion works on GPU memory space. We update the buffer offset and size once there is a new request available in inference queue $Q_I$. Therefore, the temporal fusion process contains two operations: 1) Place new request memory adjacent to current memory space; 2) Modify buffer offset and buffer accordingly. Then, when computing kernels or collective operations are called, they can operate on the exact memory space we intend, without involving in additional unnecessary memories.

\begin{algorithm}[h]
    \caption{Main stream for token generation}
    \label{alg:fusion}
    \scriptsize
    
    \begingroup
    \color{teal}
    /* Create an inference map to track every request */ \\
    \endgroup 
    \texttt{InferenceMap inmap;} \\
    
    \begingroup
    \color{magenta}
    \textbf{while not finish} /* Loop */ \\
    /* 1. Iteratively generate new tokens for current requests */ \\
    /* 2. At the start of every loop, pulling for new requests ready for token generation */ \\
    \endgroup 
    \hspace*{2em}\textbf{\textcolor{violet}{If}} {\texttt{inque.get(req)}} \textbf{\textcolor{violet}{then}}\\
    \begingroup
    \color{teal}
    \hspace*{2em}/* Kernel operations need to cover the buffer region of the new request */ \\
    \endgroup 
    \hspace*{4em}\texttt{\textcolor{blue}{Update}(offset, size);} \\
    \hspace*{4em}\texttt{inmap.insert(req);} \\
    \begingroup
    \color{violet}
    \hspace*{2em}\textbf{\textcolor{violet}{End If}} \\
    \endgroup 
    
    \begingroup
    \color{teal}
    \hspace*{2em}/* Start token generation */ \\
    \endgroup 
    \hspace*{2em}\texttt{\textcolor{blue}{cublasGemm}(ctxt\_buffer + offset, size, ...);} \\
    \hspace*{2em}\texttt{\textcolor{blue}{LayerNorm}(ln\_buffer + offset, size, ...);} \\
    \hspace*{2em}\texttt{\textcolor{blue}{GenericActivation}(act\_buffer + offset, size, ...);} \\
    \hspace*{2em}\texttt{\textcolor{blue}{NCCLAllreduce}(reduce\_buffer + offset, size, ...);} \\
    \hspace*{2em}\texttt{\textcolor{blue}{...}} \\

    \begingroup
    \color{teal}
    \hspace*{2em}/* Check if any request finishes, so that it's buffers can be evicted */ \\
    \endgroup 
    \hspace*{2em}\texttt{inmap.FindAndEvict(require\_shuffle);} \\
    
    \hspace*{2em}\textbf{\textcolor{violet}{If}} {\texttt{require\_shuffle}} \textbf{\textcolor{violet}{then}}\\
    \begingroup
    \color{teal}
    \hspace*{2em}/* Perform memory shuffle, making buffers tight and contiguous  */ \\
    \endgroup 
    \hspace*{4em}\texttt{inmap.LaunchMemShuffle();} \\
    \begingroup
    \color{violet}
    \hspace*{2em}\textbf{\textcolor{violet}{End If}} \\
    \endgroup 
    \begingroup
    \color{magenta}
    /* End loop */ \\
    \endgroup 
\end{algorithm}

\subsection{Memory shuffle for creating contiguous buffer} 
\label{shuffle}
We have discussed in preliminaries that the arrival of requests is random, however, if the inference of every request will always reach the maximum output length before it gets an EOS to finish, then the memory management would be very simple and straightforward. Assume that the first request 
is assigned \texttt{mem\_id=0}, the second request assigned \texttt{mem\_id=1}, the third request assigned \texttt{mem\_id=2}, etc., then we only need to increase \texttt{buffer\_size} when a new request arrives, and increment \texttt{buffer\_offset} when a request finishes, and the memory space is guaranteed to be contiguous (assume there is enough memory that allows us to monotonically increase \texttt{buffer\_offset}). The reason is that since all requests require fixed iterations, then the whole inference pipeline can be seen as a FIFO queue, where the request that arrives first will also evict from memory first. However, such an ideal assumption might not be true for complicated real inference scenarios.

\begin{figure}[h]\centering
\vspace{-1em}
\includegraphics[width=1.0\linewidth]{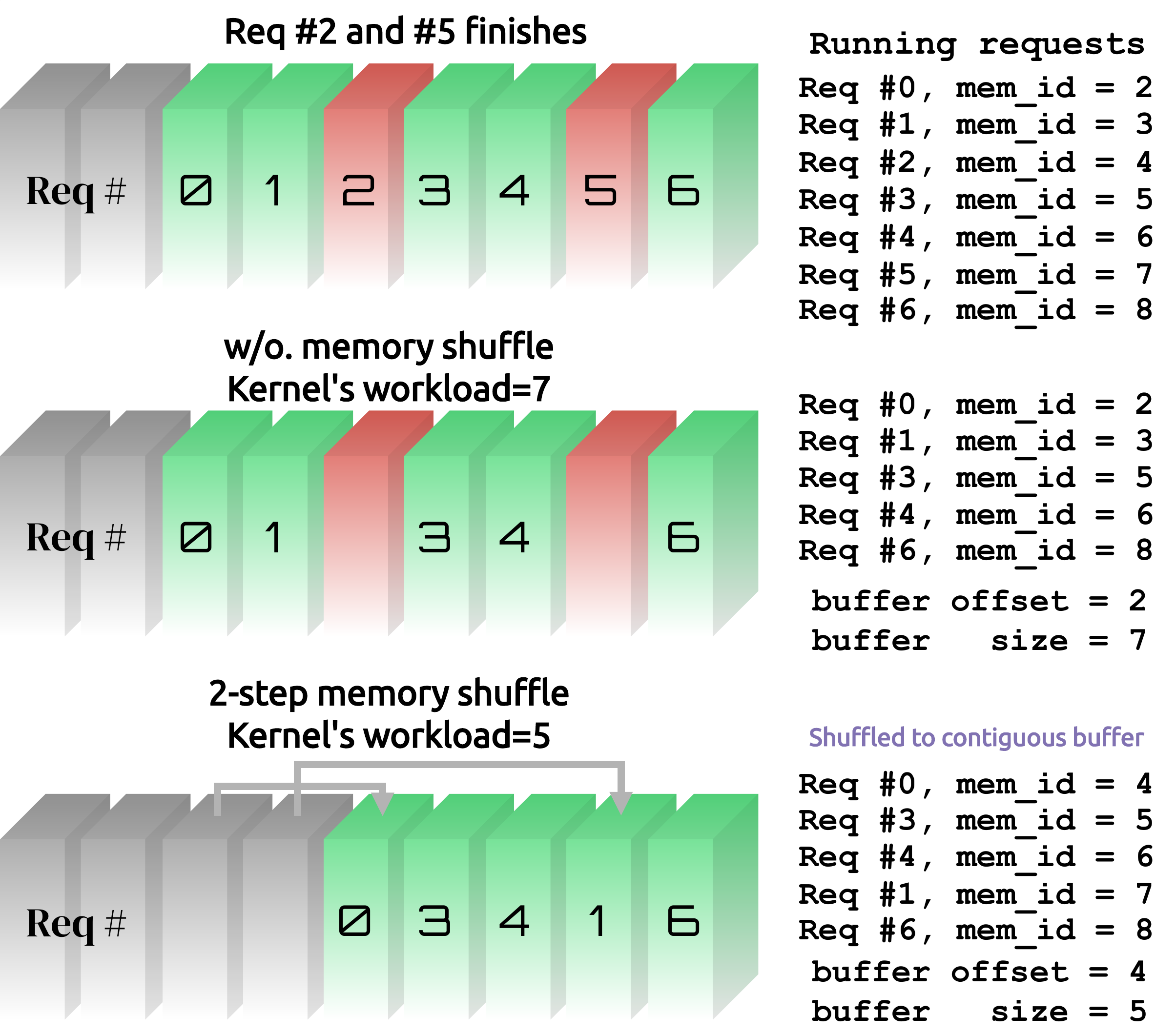}
\caption{Reordering the buffer using memory shuffle, guaranteeing a contiguous memory of running requests.}
\label{fig:memory_shuffle1}
\vspace{-1em}
\end{figure}
\begin{algorithm}[h]
    \caption{Find Shuffled Memory Region}
    \label{alg:shuffler}
    \scriptsize

    \begingroup
    \color{teal}
    /* Initialization */ \\
    \endgroup 
    \texttt{total\_cost = 0;} \\
    \texttt{non\_zero = 0;} \\

    \begingroup
    \color{teal}
    /* Loop to find non-zero elements in array and their total cost */ \\
    \endgroup 
    \textcolor{blue}{\texttt{for}} \texttt{(i = 0; i < arr.size(); ++i)} \\
    \textbf{\textcolor{violet}{If}} \texttt{arr[i] != 0} \textbf{\textcolor{violet}{then}} \\
    \hspace*{2em}\texttt{non\_zero += 1;} \\
    \hspace*{2em}\texttt{total\_cost += arr[i];} \\
    \textbf{\textcolor{violet}{End If}} \\

    \begingroup
    \color{teal}
    /* Initialize min\_cost and window\_cost */ \\
    \endgroup 
    \texttt{min\_cost = $\infty$;} \\
    \texttt{window\_cost = 0;} \\
    \texttt{mem\_offset = 0;} \\

    \begingroup
    \color{teal}
    /* Calculate window cost and update minimum cost and memory offset accordingly */ \\
    \endgroup 
    \textcolor{blue}{\texttt{for}} \texttt{(i = non\_zero; i < arr.size(); ++i)} \\
    \hspace*{2em}\texttt{window\_cost += arr[i] - arr[i - non\_zero];} \\
    \hspace*{2em}\texttt{current\_cost = total\_cost - window\_cost;} \\
    \hspace*{2em}\textbf{\textcolor{violet}{If}} \texttt{current\_cost < min\_cost} \textbf{\textcolor{violet}{then}} \\
    \hspace*{4em}\texttt{min\_cost = current\_cost;} \\
    \hspace*{4em}\texttt{mem\_offset = i - non\_zero + 1;} \\
    \hspace*{2em}\textbf{\textcolor{violet}{End If}} \\

    \begingroup
    \color{magenta}
    /* Return memory offset */ \\
    \endgroup 

\vspace{-1em}
\end{algorithm}
As we discussed before, for an generative model, inference requests are likely to differ in max output lengths. Some requests only need a few output tokens, whereas others might require thousands. More commonly, even for an inference server that has already set a max output length for all requests, the inference might output an EOS token, such as ``\$", before it reaches the length limitation. In this case, keep generating new tokens for this request will waste lots of computing power and add additional latency as any tokens following the ``\$" will be considered invalid. Thus, it is clear that when a request sees an ending token \$ or reaches the length limitation, it should immediately evict from the memory. Fig~\ref{fig:memory_shuffle1} depicts such a situation, where requests 2 and 5 finish, then after they evict, how do we manage the memory space?

If we simply keep buffer offset and size unchanged, then those evicted memories are detrimental to the inference pipeline, as both computing kernels and collective communication can only process contiguous memory buffers, and they still have to cover \texttt{mem\_id} 4 and 7. Thus, we need an efficient algorithm to shuffle and reorder the memory by moving all valid buffers together to form a new continuous memory space. The problem now becomes how to minimize the amount of memory that needs to be moved and therefore not introduce too much overhead, as the inference server will block following iterations until memory is properly managed. 

To abstract the problem, given an array of 0 and 1, where 0 denotes empty memory space, and 1 denotes valid, we need an algorithm that can group all 1 together while moving as less number of elements as possible. Here we use a sliding window algorithm~\ref{alg:shuffler} with time complexity $O(n)$ to achieve it. 

\begin{figure*}[t]\centering
    \begin{minipage}{\textwidth}
        \begin{minipage}{0.24\textwidth}
            \includegraphics[width=\linewidth]{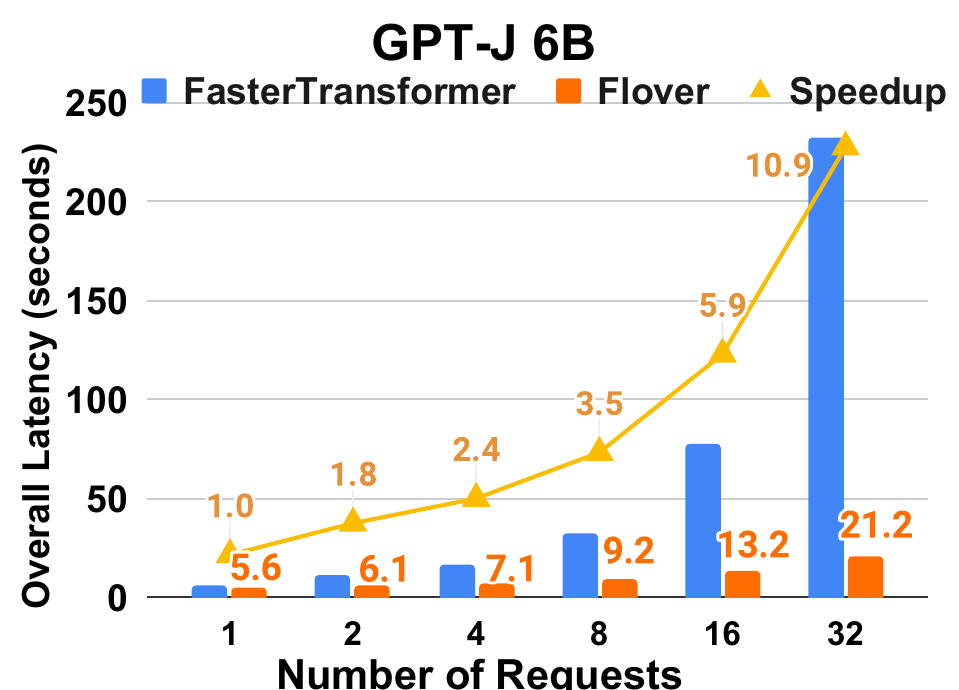}
        \end{minipage}
        \begin{minipage}{0.24\textwidth}
            \includegraphics[width=\linewidth]{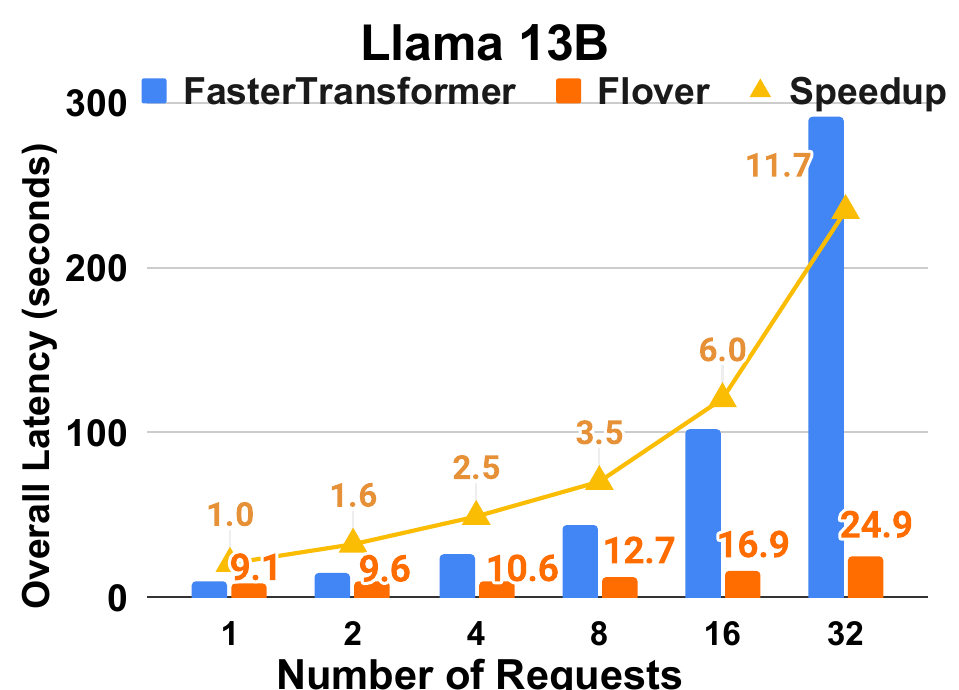}
        \end{minipage}
        \begin{minipage}{0.24\textwidth}
            \includegraphics[width=\linewidth]{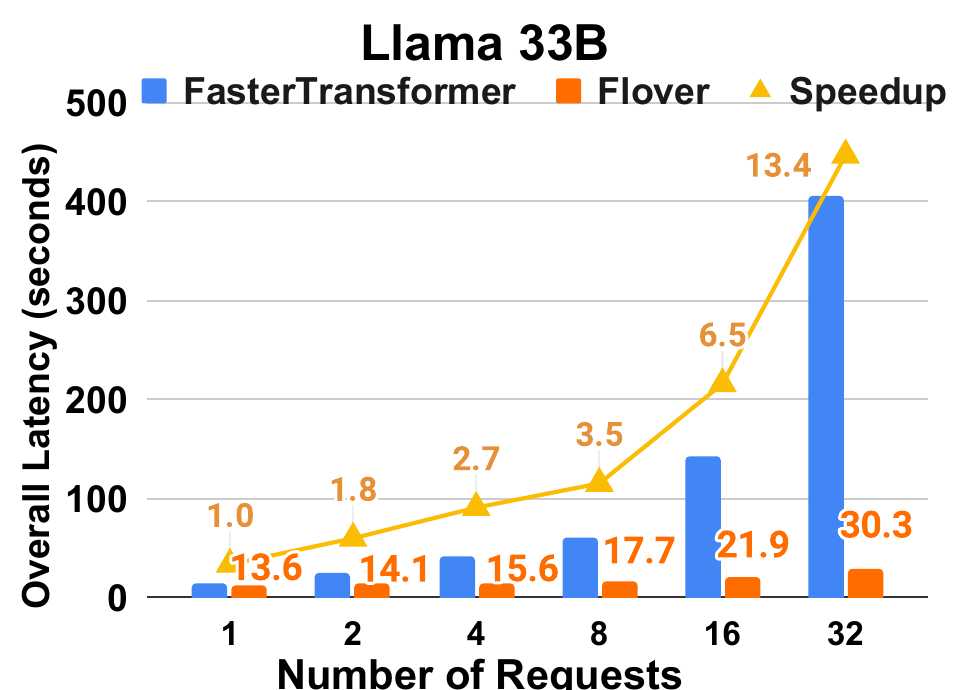}
        \end{minipage}
        \begin{minipage}{0.24\textwidth}
            \includegraphics[width=\linewidth]{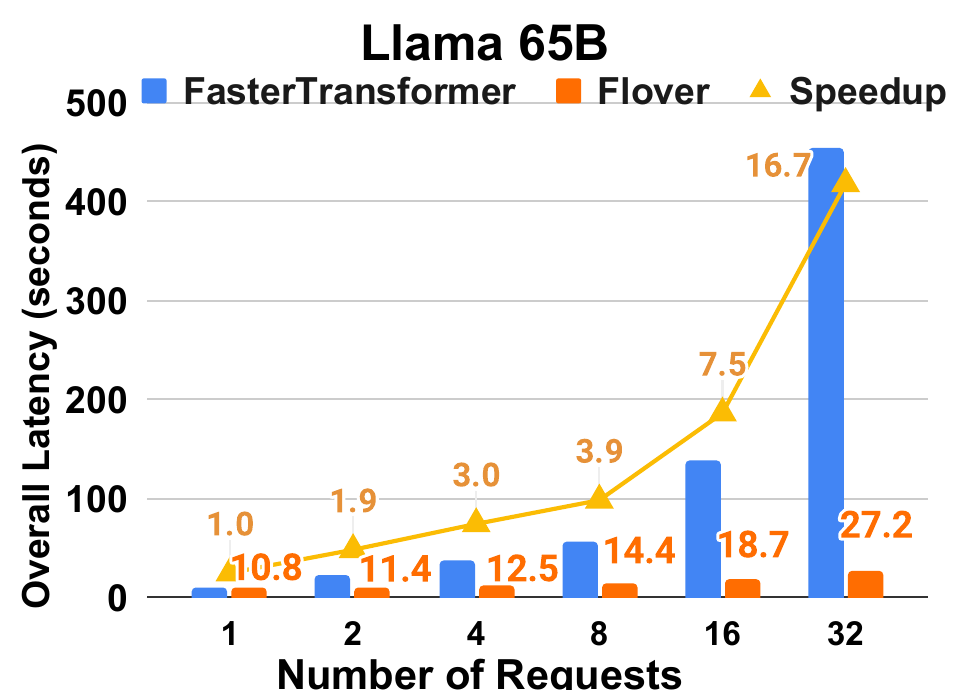}
        \end{minipage}
        \captionsetup{labelformat=empty}
        \caption*{(a). Parallel inference on different models. We measure the overall time spent on parallel inference \textbf{1, 2, 4, 8, 16, 32} requests.}
        \vspace{-0.8em}
    \end{minipage}
    
\vspace{1em}
\hspace*{-0.75cm} 
\begin{minipage}{1.05\textwidth}
    \includegraphics[width=\linewidth]{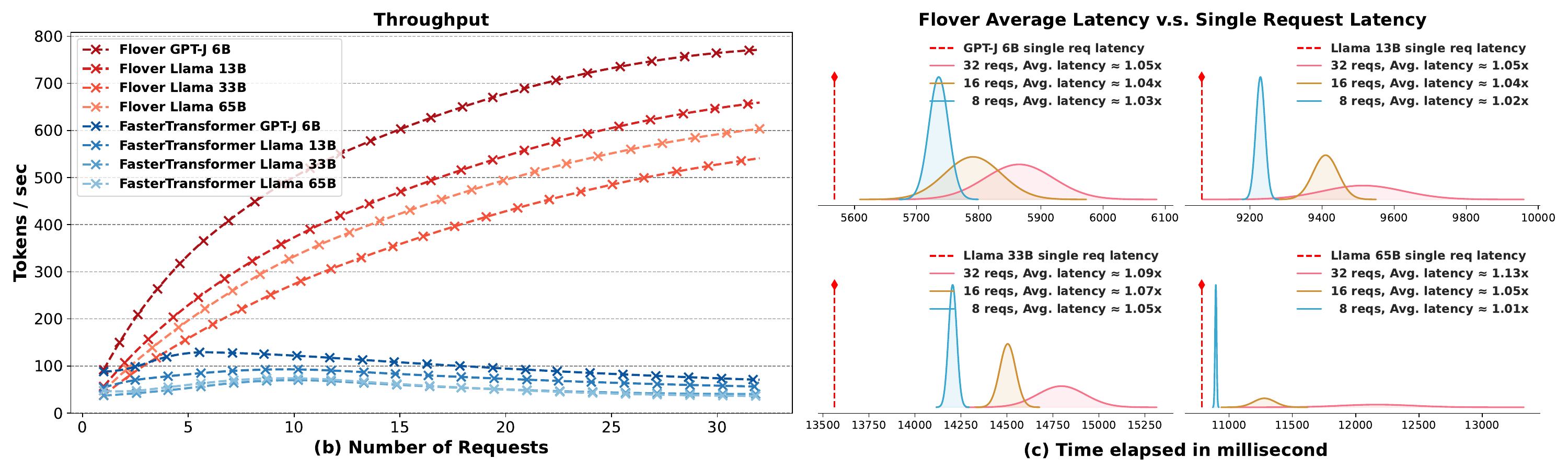}
\end{minipage}

\caption{GPT-J, Llama 13B, 33B, and 65B runs on 1, 1, 2, 4 GPUs respectively. (b) Throughput and (c) latency analysis on GPT-J and Llama. We use statistical data to model the average inference latency over the latency of inferring a single request.}
\vspace{-0.8em}
\label{fig:exp1}
\end{figure*}

Since an ideal shuffle will result in a contiguous memory region of size $n$ if there are $n$ 1's in the array. Thus we only need to locate where this memory region of size $n$ should lay, and we can copy those 1's outside of this region in. Algorithm~\ref{alg:shuffler} shows how to find the offset of this memory region. Fig~\ref{fig:memory_shuffle1} (last row) illustrates the shuffled memory region and the corresponding shuffle strategy. Note that our algorithm guarantees that the total amount of memory movement is minimized, but might disorder the memory offsets of requests. Therefore, for each request running in the inference model, it also tracks GPU memory offsets of all its tensors.



\section{Experiments}
\label{experiment}

\subsection{Setup}
As we emphasized, on both single GPU cases and distributed scenarios where other advanced parallel strategies like tensor parallel~\cite{shoeybi2019megatron} are already deployed, Flover can largely propel autoregressive model inference with its unique and efficient workflow. Therefore, we conduct thorough ablation experiments on both cases to study how Flover improves inference efficiency at a fine-grained level, and we use various analysis methods to profiling how Flover outperforms existing solutions.

\label{setup}
\textbf{Hardware:} We conduct all experiments on NVIDIA A100 80GB GPUs with AMD EPYC 7763 64-Core Processor. Each computing node has four GPUs connected by NVLINK. All collective operations are performed by the NVIDIA Collective Communications Library~\cite{nccl} (NCCL).


\textbf{Software:} We implement our Flover framework based on NVIDIA FasterTransformer~\cite{ft} C++ codebase, which is one of the most widely used Triton~\cite{Triton} backends and large language model (LLM) solutions. For the following experiments, we use the following famous language models --- GPT-J~\cite{gpt-j} 6B, Llama~\cite{touvron2023llama} 6B, Llama~\cite{touvron2023llama} 13B, Llama~\cite{touvron2023llama} 33B, and Llama~\cite{touvron2023llama} 65B. GPT-J~\cite{gpt-j} is created by EleutherAI, a community-driven organization that aims to promote open-source AI research. It has 6 billion parameters and was trained on The Pile~\cite{gao2020pile}. Llama~\cite{touvron2023llama} are the state-of-the-art foundation language models released by Meta AI, which are trained on 1.4T tokens of CommonCrawl, Github, Wikipedia, ArXiv, etc. 

We compare our results mainly to FasterTransformer, as ORCA~\cite{yu2022orca} is not open-sourced for evaluation.

\setcounter{figure}{2} 
\begin{figure*}[t]\centering
\small
    \begin{minipage}{.52\textwidth}
        \includegraphics[width=\linewidth]{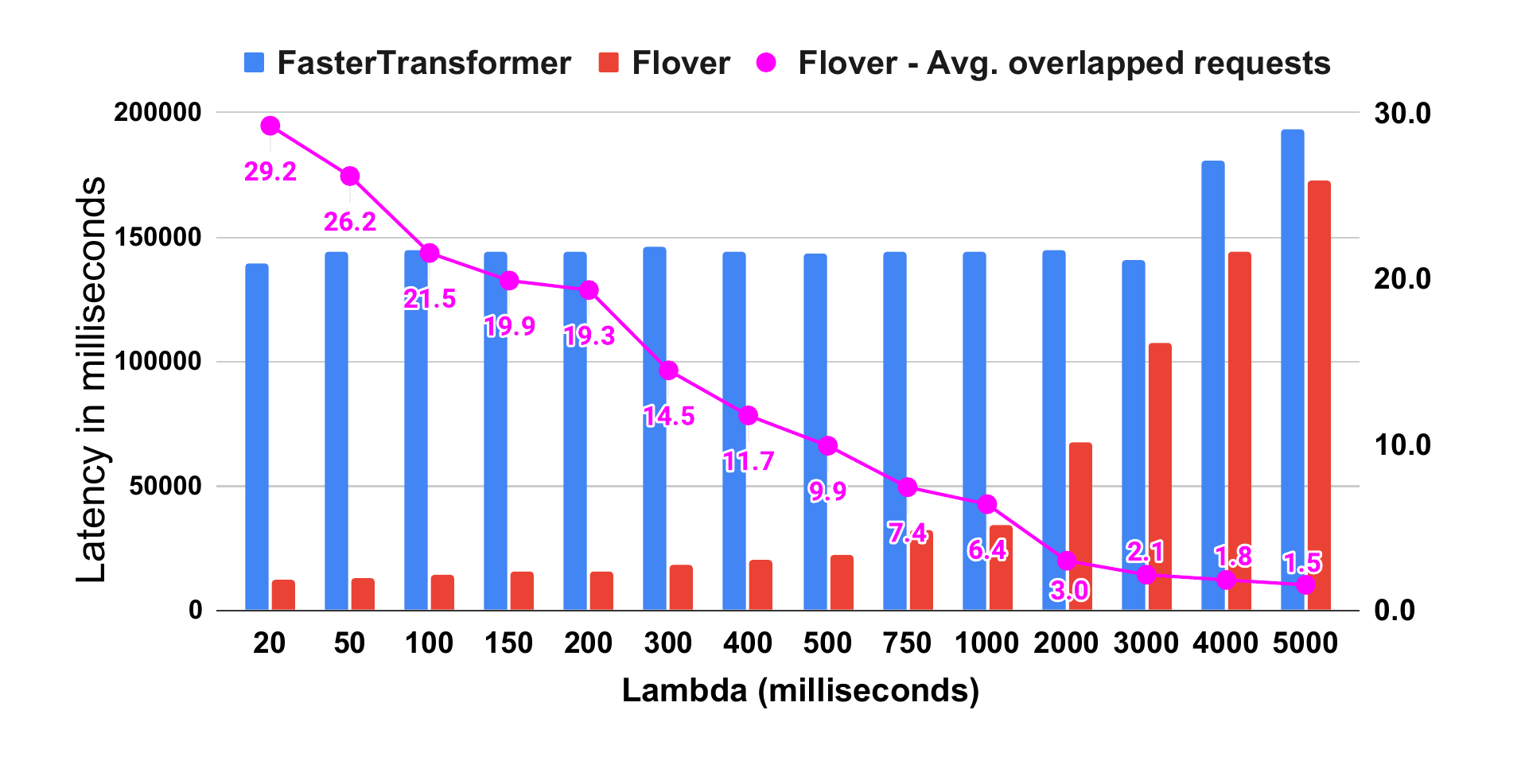}
        \captionsetup{labelformat=empty,labelsep=none}
        \vspace{-3em}
        \caption{(a)}
        
    \end{minipage}
    \hspace{0.01\textwidth}
    \begin{minipage}{.45\textwidth}
    \vspace{1em}
    \setlength{\tabcolsep}{1pt}
    \renewcommand{\arraystretch}{0.98}
        \begin{tabular}{c|c| c | c |c |c|c|c}
            \toprule[2pt]
             $\lambda$ (ms) & 20 & 50 & 100 & 150 & 200 & 300 & 400  \\
            \toprule[1pt]
            Total Iters. & 557 & 616  &  748 & 888 & 911 & 1174 & 1366\\
            \hline
            Overlap & 91.3\% & 81.8\% & 67.3\% & 62.1\% & 60.3\% & 45.2\% & 36.7\%\\
            \hline
             Speedup & 11.2x & 11.1x & 10.2x & 9.3x & 9.1x & 7.9x & 7.1x\\
            \hline
            
            \toprule[2pt]
            $\lambda$ (ms) &  500 & 750 & 1000 & 2000 & 3000 & 4000 & 5000 \\
            \toprule[1pt]
            
            Total Iters. & 1537 & 2239 & 2621 & 5483 & 7738 & 8902 & 10694 \\
            \hline
            Overlap & 31.0\% & 23.2\% & 20.0\% & 9.3\% & 6.7\% & 5.7\% & 4.8\%\\
            \hline
             Speedup &  6.5x & 4.5x & 4.2x & 2.1x & 1.3x & 1.3x & 1.1x \\
            \hline
            \toprule[2pt]
        \end{tabular}
        \vspace{0.5em}
        \captionsetup{labelformat=empty,labelsep=none}
        \caption{(b)}
    \renewcommand{\arraystretch}{1}
    \end{minipage}
    \vspace{-0.4em}
\caption{Total latency of inferring 32 requests follow the Poisson process. Time intervals between requests are randomly sampled from the exponential distribution with different $\lambda$. A single request takes 5800ms on the inference server. The right-side table further shows the number of total iterations for inferring 32 requests, requests overlap, and speedup to FasterTransformer, following the Poisson process of different $\lambda$. A single request requires 512 iterations to generate all output tokens.}
    \vspace{-1em}
    \label{fig:poisson1}
\end{figure*}

\subsection{Temporal fusion with constant time interval}
In this section, we start with analyzing how efficient Flover is when using temporal fusion to process multiple requests in parallel. As discussed, the real case of arrivals of requests is considered a Poisson process, where the time interval between two requests is a random variable from the exponential distribution. However, for simplicity, in this part, we will use a constant time interval of 500ms to study the parallel efficiency, as this is also adopted by some inference frameworks. Notice that for all models, the average inference latency for a single request is much longer than 500ms, therefore it leaves great potential for parallel acceleration. 

Fig~\ref{fig:exp1} (a) compares Flover to FasterTransformer on four different models. For GPT-J 6B and Llama 13B, we run on 1 GPU without tensor parallelism. For Llama 33B, we run on 2 GPUs with tensor parallelism of size 2. And for Llama 65B, we use 4 GPUs to perform degree-4 tensor parallelism. When we only have 1 request running for inference, both frameworks deliver similar performance on all models as there is no difference in the workflow. When increasing the parallel requests to 2, a salient disparity in latency performance is observed. Flover provides an average \textbf{1.8x} speedup in overall latency. This is largely due to Flover only initiating half of kernel calls and collective operations, while in FasterTransformer, multiple model instances are competing for resources, leading to significant overhead in context switching. Therefore, Flover delivers increasingly higher parallel efficiency when we keep increasing the concurrency of  requests. At 32 concurrency, Flover only spends $9\%$ of time to complete all requests' inference on GPT-J 6B model and $6\%$ on Llama 65B. And as we involve higher degree of tensor parallelism, additional NCCL collective calls further slow down FasterTransformer. In Llama 65B model, Flover provides \textbf{16.7x} speedup against the baseline.

Fig~\ref{fig:exp1} (b) further analyzes the throughput of both frameworks on these models. As Flover fuses the inference of multiple requests by enabling individual kernels to operate on larger contiguous buffer pieces, which is similar to increasing the batch size from 1 to 32, it achieves optimal utilization of GPU resources or an embarrassing parallel. \label{context_switching}While in FasterTransformer, since each kernel is still operating on a single request batch, increasing concurrency cannot benefit the throughput, instead, because of frequent launching and context switching, we observe a deterioration when launching too many kernels at 32 parallel requests. Fig~\ref{fig:exp1} (c) provides a more detailed analysis of the average latency of each request, as this directly affects the user experience of their request. As the framework is dealing with different number of parallel requests at any time step, so we present a statistical result on the latency. With 8 parallel requests, the average latency of each request is only \textbf{3\%} longer than solely inference such a request. When we increase the concurrency to 32, it takes slightly longer time for delivering the generation, at around \textbf{8\%} slower than solely inference such a request. The behavior is expected as we increase the workload of kernels, though the overall throughput gets better, the average latency per request will increase as well. For FasterTransformer however, due to context switching, we observe that almost all requests are finished together at the very end, leading to extremely worse latency performance.

\subsection{Temporal fusion with Poisson process}
Consider such a request $r_i$, containing an inference task that takes the inference server about $t_{r}$ to finish. Let's denote the time interval between request $r_0$ and $r_1$ as $\tau$. If $\tau \ll t_r$, then most of the time, $r_0$ and $r_1$ are temporally overlapped in the inference server. If $\tau \gtrsim t_r$, then requests are considered sequentially processed. In practice, however, overlapping two requests might sightly affect $t_r$ as we stated previously, by $1\%$ to $13\%$ as shown in Fig~\ref{fig:exp1} (c). Here we stick to it as it is enough for our analysis.

As we discussed, the arrival times of inference requests are not fixed or predictable in a strict sense. Instead of adhering to a constant time window or a constant interval between the arrival of each request, the process can be modeled as a Poisson process, in which the exponential distribution models the varying time intervals between the arrivals of requests. Here each request is with a 512 output tokens limitation. Bars in Fig~\ref{fig:poisson1} compare the total inference latency on 32 requests using FasterTransformer~\cite{ft} and Flover respectively, under a span of $\lambda$ in $[20ms, 5000ms]$. The yellow line reports the average number of overlapped requests in the overall inference, which is in inverse proportion to $\lambda$. When $\tau=20$ms, almost all requests are parallel processed by the inference server, while when $\tau=5000$ms, on average only 1 or 2 requests can temporally overlap with each other. Table in Fig~\ref{fig:poisson1} provides a more detailed stat on the Poisson process. $Overlap$ is dividing the average number of temporally overlapped requests by the total number of requests. Total Iters. counts from the first request's output token to the end token of the last request. Given that one request requires 512 iterations for inference, the larger the overlap, the more performance gain Flover can provide, as it is able to optimize most computing and communication during the inference. Also noteworthy is that in concurrent model instances, the time interval does not dominate the overall latency until it reaches 4000ms. This is due to operating multiple instances which introduce too much overhead for the inference server as we will analyze in~\ref{profiling}, resulting in severe degradation in performance. 


\subsection{Memory shuffle for non-uniform requests}

We have so far analyzed different arrival patterns of requests, e.g. constant, random. However, in real-world scenarios, requests from various users might vary drastically in the total number of iterations, which is another random variable. The distribution of the total number of iterations (i.e., the length of the generated sequences) before an end-of-sequence (EOS) token appears in a sequence generated by an autoregressive model like GPT~\cite{brown2020language,gpt-j,https://doi.org/10.48550/arxiv.2204.06745,openai2023gpt4,radford2019language, gpt-neo} largely depends on the specifics of the model and its training data. If the model has been trained on a dataset where text sequences typically have a certain length, it will likely generate sequences of similar length when run on similar data. Moreover, the generation process in autoregressive models inherently includes a degree of randomness. This randomness can cause variability in the length of the generated sequences, making it hard to fit a simple distribution. And techniques such as beam search, top-k sampling, or temperature adjustments 
\begin{figure}[h]\centering
    \vspace{-1em} 
        \includegraphics[width=.45\textwidth]{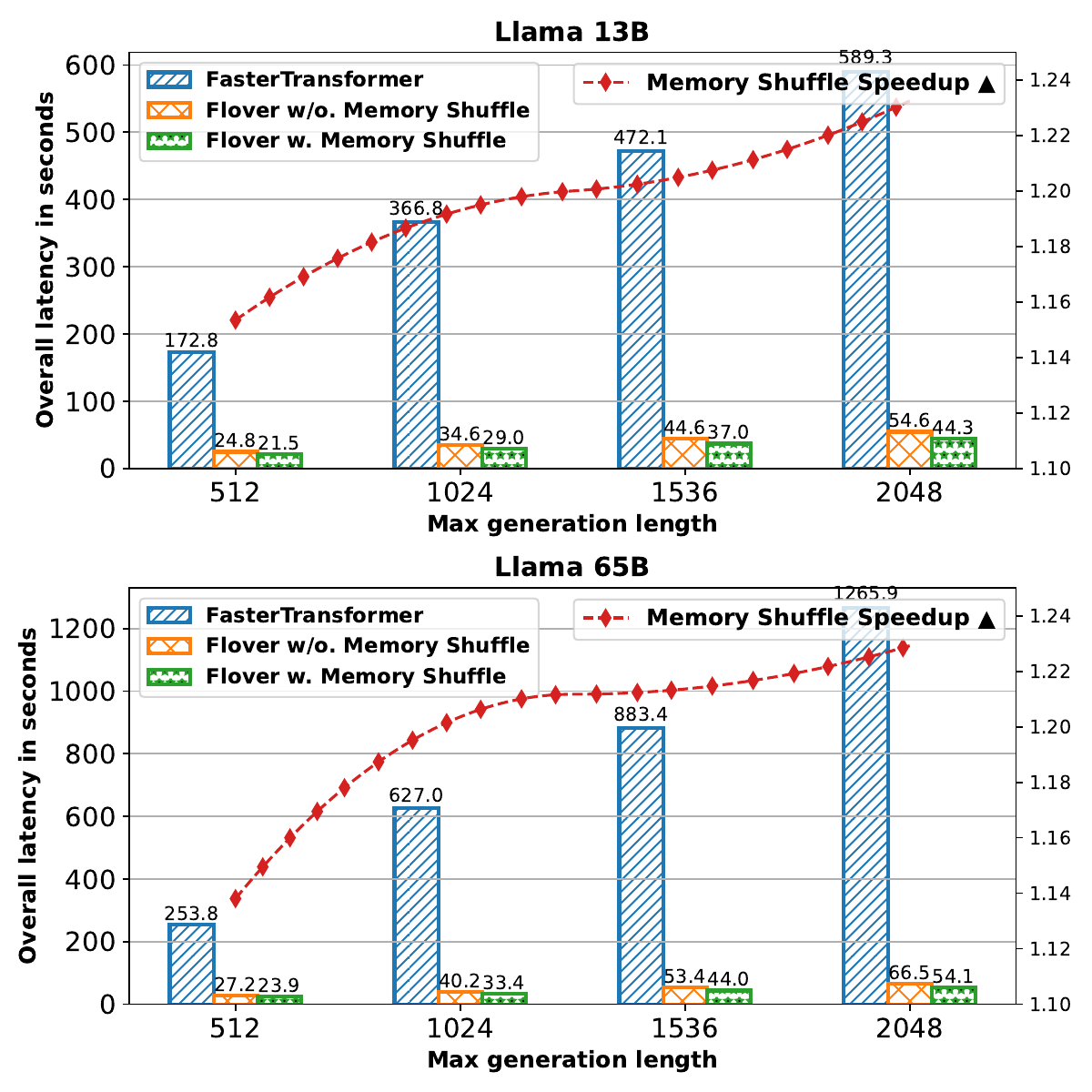}       
\caption{Total latency of inferring 32 requests with random numbers of token generation. Requests' generation length is sampled from a uniform distribution with a lower bound fixing at 128, and the upper bound varies on the x-axis.}
    \vspace{-1em}
\label{fig:shuffle_res1}
\end{figure}
used during the generation process can also affect the length of the output sequences. Given these factors, to better study how different frameworks perform in the most-uncertain scenarios or worst-case, we adopt a uniform distribution $U_l(a,b)$ to model and sample requests' total number of iterations, where all values are equally distributed.

In this experiment, we will vary $b$, to mimic the use cases of Flover in various autoregressive models. We set the number of requests to 32 to approach the real distribution and reduce variance.
 In Fig~\ref{fig:shuffle_res1}, we compare our method with the baseline FasterTransformer. Notice that Flover without memory shuffle refers to the naive solution we showed in Fig~\ref{fig:exp1}, which will not perform any memory shuffle operations but leave those finished requests' buffers within the contiguous memory space. It is clear that when enabling memory shuffle after requests evict from the compute stream, Flover is able to gain more performance during the inference, as memory shuffle will reorder the buffer to make sure evicted ones are no longer part of the computation. Also noteworthy is that, for $U_l$ on the interval $[a,b]$, the standard deviation is given by the $\sigma=\sqrt{\frac{(b-a)^2}{12}}$. Therefore, as we increase the upper bound of $U_l$, requests tend to have more various numbers of iterations, which means there will be more evicted buffers as requests finish. Averagely, memory shuffle delivers $20\%$ improvement in latency compared to our vanilla design. While in total, it delivers a \textbf{23.4x} speedup in overall inference latency against the baseline.

\subsection{Profiling on hardware scheduling patterns}
\label{profiling}
We've mentioned in early sections that compared to existing parallel inference strategies, Flover provides both scalability and instantaneity in dealing with heavy load scenarios. Fig~\ref{fig:progress} shows the cumulative token generation progress of processing 32 requests in parallel. We explicitly present 4 settings of dynamic batching strategy and also compare them to the concurrent instances strategy which is used by FasterTransformer. Although let dynamic batching wait for a long window, e.g. 16000ms, will result in higher throughput in token generation, 
\begin{figure}[h]\centering
\vspace{-1em}      
        \includegraphics[width=.48\textwidth]{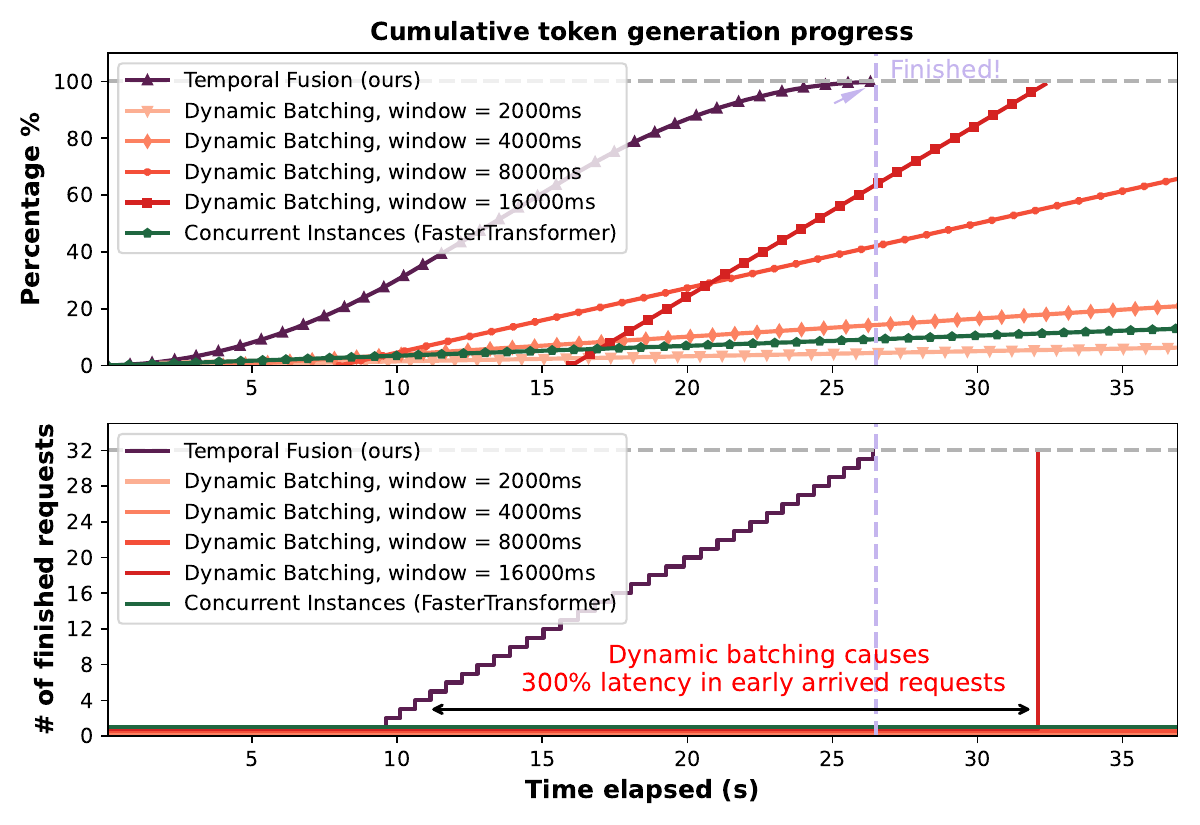}
          
\caption{Cumulative progress in inference 32 requests with different parallel strategies. Results are conducted on Llama 13B with a maximum generation length of 512. Requests are sent to the server with a fixed time interval of 500ms.}
    \vspace{-1em}
\label{fig:progress}
\end{figure}
it also introduces severe latency, especially to the early arrived requests. As shown in the second figure, dynamic batching largely delays all requests in the same window by $300\%$ compared to the temporal fusion strategy that Flover adopts.

In Fig~\ref{fig:tp}, we further analyze the hardware scheduling pattern of the concurrent instances strategy used by FasterTransformer. As we have provided theoretical explanations in~\ref{context_switching}, we directly dive into the profiling results. In the Nsight interface, blue blocks represent the CUDA kernel calls, while the CPU threads status is shown at the bottom. In FasterTransformer (Fig~\ref{fig:tp} (a)), since each model instance has to run by a dedicated thread, parallel inference on 32 requests will require launching that many threads. The CUDA profiling clearly shows that the randomness and chaos due to each instance trying to launch separate kernels and the CPU also needs to handle context switching caused potentially by over-subscription. In Flover example, the CUDA profiling is very clear and uniform, as there is only 1 thread and 1 instance issuing kernel calls. Notice that each segment here represents one iteration in the model inference, while this streamlined pattern cannot be observed in FasterTransformer. As we demonstrated in Fig~\ref{fig:pipeline}, Flover only keeps two threads spinning during the entire runtime. This is also shown in Fig~\ref{fig:tp} (b) bottom part.



\section{Acknowledgement} 
We thanks contributors in~\cite{llama_support} for providing the Llama implementations with FasterTransformer.
\section{Conclusions}
\label{conclusions}
\begin{figure*}[t]\centering
\vspace{-1em}
\begin{subfigure}{1\textwidth}
\centering
    \includegraphics[width=0.96\linewidth]{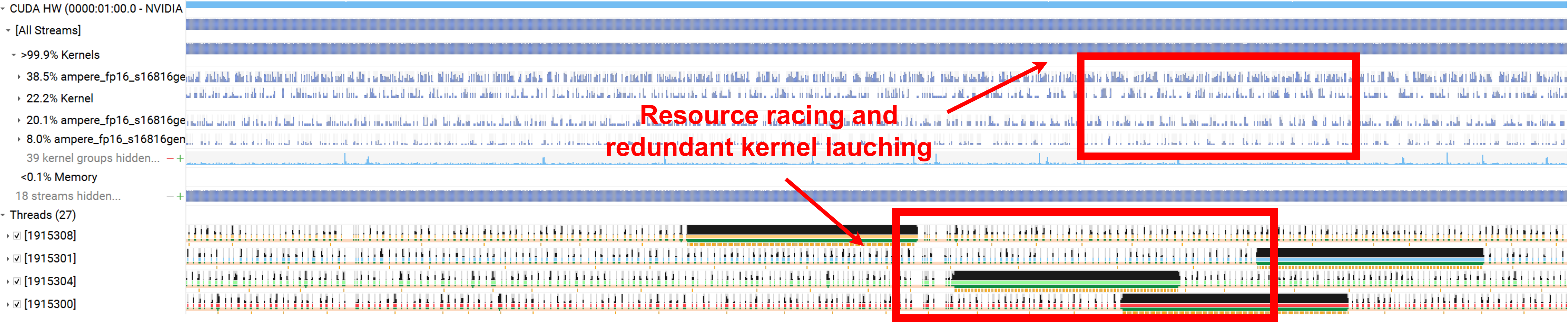}
\vspace{-0.5em}
\caption{\normalsize NVIDIA Nsight Profiling on baseline FasterTransformer. All model instances compete for resources.}
\end{subfigure}
    
    \begin{subfigure}{1\textwidth}
\centering
    \includegraphics[width=0.96\linewidth]{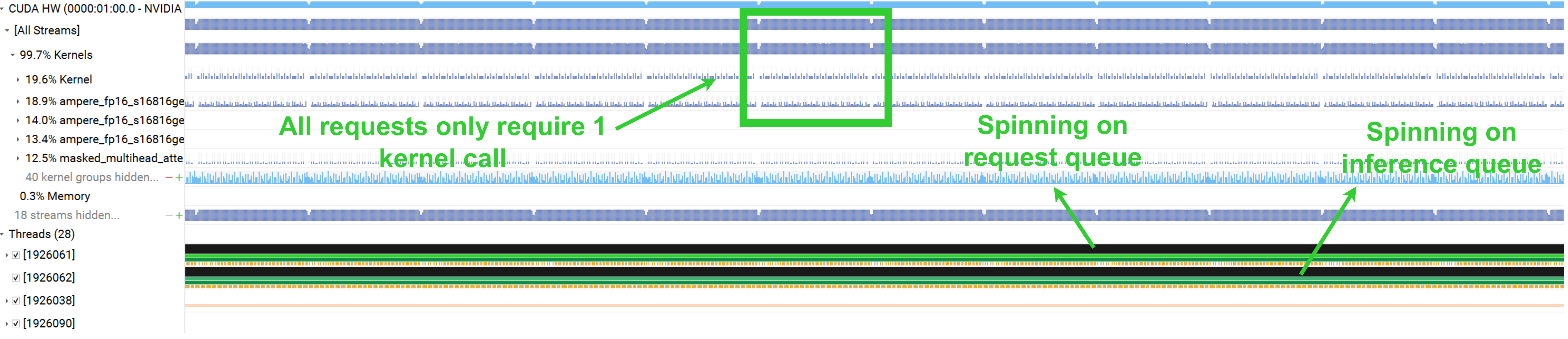}
\end{subfigure}
\vspace{-1.8em}
\caption{(b) NVIDIA Nsight Profiling on Flover. Each green box denotes parallel generating 1 token for all requests.}
\vspace{-1em}
\label{fig:tp}
\end{figure*}
We have proposed a novel temporal fusion framework (Flover) for efficient autoregressive model inference across various industrial scenarios. Unlike existing solutions that either require a delayed batching of requests or launch multiple model instances to serve the need, which lacks flexibility and causes severe overhead in response time, Flover innovatively leverages temporal parallelism of autoregressive models, providing instantaneous inference on incoming requests while being able to seamlessly fuse new requests to proceeding ones regardless of their temporal gaps. By employing an efficient memory shuffle algorithm, our solution enhances hardware utilization and substantially reduces the overhead in computing and communication, guaranteeing a highly efficient and performant inference framework. Being synergistically coalesced with the advanced tensor parallel technique, Flover achieves optimal management on both single GPU and distributed inference scenarios, ensuring robustness and scalability in diverse autoregressive model inference landscapes. We hope that this work sparks further research and innovations, fostering new methods and techniques that build upon this foundation.

\section{Related work} 
A few works have investigated accelerating the inference of auto-regressive generative models. ORCA~\cite{yu2022orca} provides a similar iteration batching mechanism, however, their solutions can only deal with uniform requests with fixed generation length and hence do not support memory eviction of early finished requests. As we stated in our ablation experiments, we believe that real-world requests are all various and follow random arriving patterns which need to be thoroughly analyzed.
\bibliographystyle{plain}
\bibliography{main}
\end{document}